\documentclass[11pt,a4paper]{article}

\usepackage{jinstpub}
\usepackage{lineno}

\usepackage{siunitx}
\usepackage{makecell}
\usepackage{comment}

\title{
\boldmath
Simulation of the CYGNO Gaseous TPC Optical Readout
}

\author[1]{F.D. Amaro}
\author[2,3]{R. Antonietti}
\author[4,5]{E. Baracchini}
\author[6]{L. Benussi}
\author[6]{S. Bianco}
\author[6]{C. Capoccia}
\author[6,7]{M. Caponero}
\author[8]{L.G.M de Carvalho}
\author[9,10]{G. Cavoto}
\author[6]{I.A. Costa}
\author[6]{A. Croce}
\author[4,5]{M. D'Astolfo}
\author[10]{G. D'Imperio}
\author[6]{G. Dho}
\author[10]{E. Di Marco}
\author[1]{J.M.F. dos Santos}
\author[4,5]{D. Fiorina}
\author[10]{F. Iacoangeli}
\author[4,5]{Z. Islam}
\author[11]{E. Kemp}
\author[4,5]{H. P. Lima Jr}
\author[6]{G. Maccarrone}
\author[1]{R.D.P. Mano}
\author[4,5]{D. J. G. Marques}
\author[6]{G. Mazzitelli}
\author[2,3]{P. Meloni}
\author[9,10]{A. Messina}
\author[1]{C.M.B. Monteiro}
\author[8]{R.A. Nobrega}
\author[8]{I.F. Pains}
\author[6]{E. Paoletti}
\author[2,3]{F. Petrucci}
\author[4,5]{S. Piacentini}
\author[6]{D. Pierluigi}
\author[10]{D. Pinci}
\author[10]{F. Renga}
\author[6]{A. Russo}
\author[6,12]{G. Saviano}
\author[1]{P.A.O.C. Silva}
\author[13]{N. J. C. Spooner}
\author[6]{R. Tesauro}
\author[6]{S. Tomassini}
\author[4,5]{S. Torelli}
\author[9,10]{D. Tozzi}

\affiliation[1]{LIBPhys; Department of Physics; University of Coimbra; 3004-516 Coimbra; Portugal}
\affiliation[2]{Dipartimento di Matematica e Fisica; Universit\`a Roma TRE; 00146; Roma; Italy}
\affiliation[3]{Istituto Nazionale di Fisica Nucleare; Sezione di Roma Tre; 00146; Rome; Italy}
\affiliation[4]{Gran Sasso Science Institute; 67100; L'Aquila; Italy}
\affiliation[5]{Istituto Nazionale di Fisica Nucleare; Laboratori Nazionali del Gran Sasso; 67100; Assergi; Italy}
\affiliation[6]{Istituto Nazionale di Fisica Nucleare; Laboratori Nazionali di Frascati; 00044; Frascati; Italy}
\affiliation[7]{ENEA Centro Ricerche Frascati; 00044; Frascati; Italy}
\affiliation[8]{Universidade Federal de Juiz de Fora; Faculdade de Engenharia; 36036-900; Juiz de Fora; MG; Brasil}
\affiliation[9]{Dipartimento di Fisica; Universit\`a di Roma Sapienza; 00185; Roma; Italy}
\affiliation[10]{Istituto Nazionale di Fisica Nucleare; Sezione di Roma; 00185; Roma; Italy}
\affiliation[11]{Universidade Estadual de Campinas  - UNICAMP;  Campinas 13083-859; SP; Brazil}
\affiliation[12]{Dipartimento di Ingegneria Chimica; Materiali e Ambiente; Sapienza Universit\`a di Roma; 00185; Roma; Italy}
\affiliation[13]{Department of Physics and Astronomy; University of Sheffield; Sheffield; S3 7RH; UK}

\emailAdd{fabrizio.petrucci@uniroma3.it}

\abstract{
Gaseous Time Projection Chambers with Optical Readout are sensitive detectors suitable for 3D measurement of low-energy particles (of order 1~keV) and are proposed for detecting rare events such as Dark Matter particle interactions. The CYGNO collaboration is developing such a detector with a high spatial and energy resolution, leveraging an innovative optical readout system. A reliable simulation of the detector response is needed to properly assess the physics reach of this technique and to better understand the performance of the detector in the development phase. Such a simulation cannot entirely rely on existing software packages; indeed, none of the available tools is capable of properly and reliably treating the different phenomena occurring in the detector, from the primary interaction in the gas volume throughout the whole detector response model, including charge transport, light production and propagation, and the response of the optical sensors. In this paper, we present a modeling of the detector response tuned on the CYGNO Optical TPC case; a description of the method is reported together with comparisons with experimental data from the LIME prototype to demonstrate the simulation performances.
}


\begin{document}
\maketitle

\newpage

\section{Introduction}
In the search for Dark Matter (DM) WIMP-like particle candidates, large regions of the spin independent particle-nucleon scattering cross section versus particle mass plane have been excluded, namely for large masses. In particular, the mass range above 10~GeV/c$^2$ has been extensively studied over the last decades. This has resulted in a compelling necessity to push the searches to lower cross sections and lower masses that are still theoretically well motivated \cite{lowM_DM1, lowM_DM2, lowM_DM3}.
To this extent, the development of new technologies to efficiently detect and reconstruct nuclear recoil (NR) events, induced by collisions of such DM particles, with an energy of order 1~keV, and to distinguish them from electron recoil (ER) events, has become more relevant \cite{lowM_DM_detection}.
Furthermore, the possibility of studying the direction of the incoming particle would be decisive in case of a positive signal to unambiguously identify it as a DM signal and to study some of its properties \cite{directionalDM}. 

A promising solution consists of a Time Projection Chamber (TPC) with optical readout
, taking advantage of the detailed event topology reconstruction capability provided by the TPC system, coupled with the high sensitivity and granularity of the latest generation light sensors. 

A reliable and detailed simulation of the response of such a detector is crucial for assessing its physics potential and for optimizing its performance, particularly during the R\&D phase.
While several simulation software packages are widely used in HEP applications, none can reliably and efficiently model the full detector response. In any case, a dedicated framework is required to accurately describe light production, propagation, and the response of the optical sensors.

In this paper, a modeling of the detector response of the CYGNO TPC is presented; data taken with a detector prototype are also used to tune the simulation and to demonstrate its reliability.   

\section{Detector overview}
\label{geant4_sec}
The CYGNO proposal is based on a TPC with optical readout \cite{CYGNOpaper}. The gas volume is filled with a mixture of He and CF$_4$ at a 60:40 ratio at atmospheric pressure and room temperature, chosen as the best compromise for our application, balancing the light yield and the detector electrical stability \cite{lemon5}.
An ionizing particle crossing the sensitive volume of the detector produces clusters of electron-ion pairs in the gas along its trajectory. The electrons are transported to the anode by a uniform electric field, maintaining the track shape. At the anode, an amplification system consisting of a three Gas Electron Multiplier (GEM)~\cite{gem0} stack amplifies the electrons through an avalanche mechanism. 
During the avalanche, the electrons ionize and excite the CF$_4$, which emits light upon its de-excitation \cite{gem1, gem2, gem3, gem4}. The light is finally collected by scientific CMOS (sCMOS) cameras and PhotoMultiplier Tubes (PMT) \cite{orange2} positioned outside the gas vessel. By combining the 2D information (x-y) with the high granularity of the sCMOS, and the fast response of the PMT in sampling the track in the third coordinate (z), it is possible to perform a 3D reconstruction of the tracks. 

\begin{figure}
    \centering
    \includegraphics[scale=0.6]{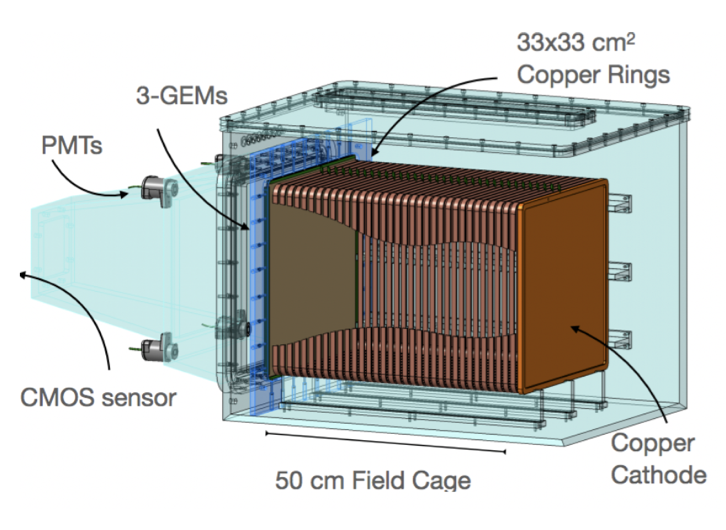}
    \includegraphics[scale=0.35]{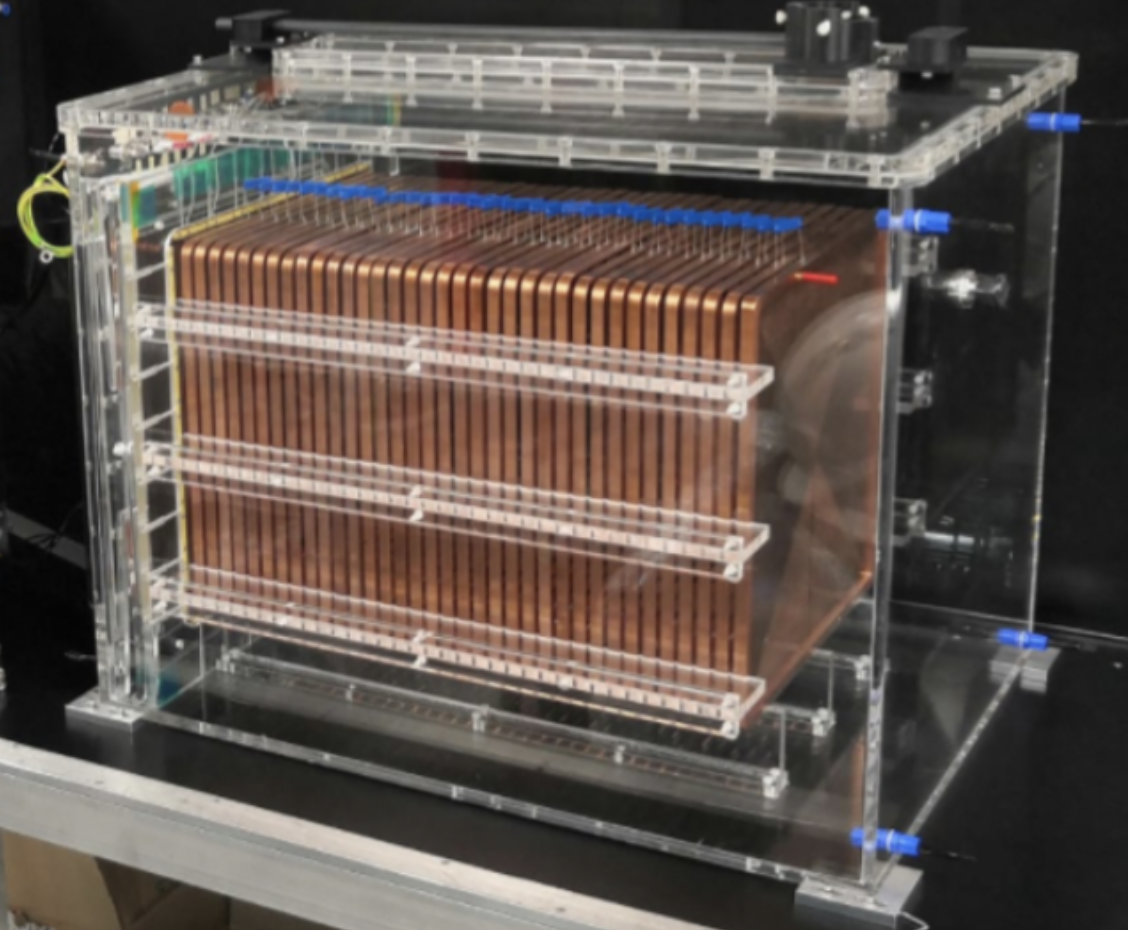}
    \caption{Left: Scheme of the LIME detector. Right: picture of the LIME detector. The field cage structure and the cathode can be seen inside the plexiglass vessel. On the upper face of the vessel the window and the trolley for the radioactive sources are clearly visible.}
    \label{fig:LIME}
\end{figure}

The latest prototype developed by our collaboration is the Large Imaging Module (LIME) \cite{limeLNF}. 
LIME has a drift length of \SI{50}{\centi\metre} and a readout area of 33×33~cm$^2$, for a total volume of about 50 liters. The electric field in the TPC is controlled by a field cage composed of 35 copper rings, which keep the electric field uniform throughout the detection volume with a value of $\sim$0.9~kV/cm, corresponding to a drift velocity of $\sim$6~cm/$\mu$s in the used gas mixture. The amplification system consists of 3 GEMs covering the total readout area. 
The GEMs have 70~$\mu$m diameter holes with a 140~$\mu$m pitch and the lower electrodes are divided in 11 high-voltage independent longitudinal sectors to increase the stability of the detector.
The light produced during the amplification process is read out by an Orca Fusion \cite{Hamamatsu2020} sCMOS camera from Hamamatsu and 4 PMTs placed along the readout plane diagonals. The camera has 2304$\times$2304~pixels with a 6.5~$\mu$m size which projects to 155~$\mu$m on the GEM plane with our optics (see Section~\ref{sec:optics}). 
The upper face of the plexiglass gas vessel includes a 5 cm wide and 50 cm long thin window sealed by a 150~$\mu$m thick ethylene-tetrafluoroethylene (ETFE) layer. This allows low energy photons (down to the keV energy) to enter the gas volume from external artificial radioactive sources used for calibration purposes. These sources can be moved back and forth by a controllable trolley holding the sources and moving along the window in the z direction from 5 to 45 cm far from the GEM. 
A scheme of the LIME detector and a picture of the actual detector field are shown in Figure~\ref{fig:LIME}.

The LIME prototype was operated at the Laboratori Nazionali di Frascati of INFN (LNF) for a long time to test the long-term stability and to study the detector response in terms of linearity and energy resolution with the use of different radioactive sources. In particular, the prototype was calibrated using various X-ray sources inducing ERs in the sensitive volume, as detailed in Section~\ref{data_samples}; these calibration data were used to tune and validate the detector simulation as described in the following sections.  

\section{Modeling the detector response}
\label{simulation}
\begin{figure}[ht]
 \centering
  {\includegraphics[width=120mm]{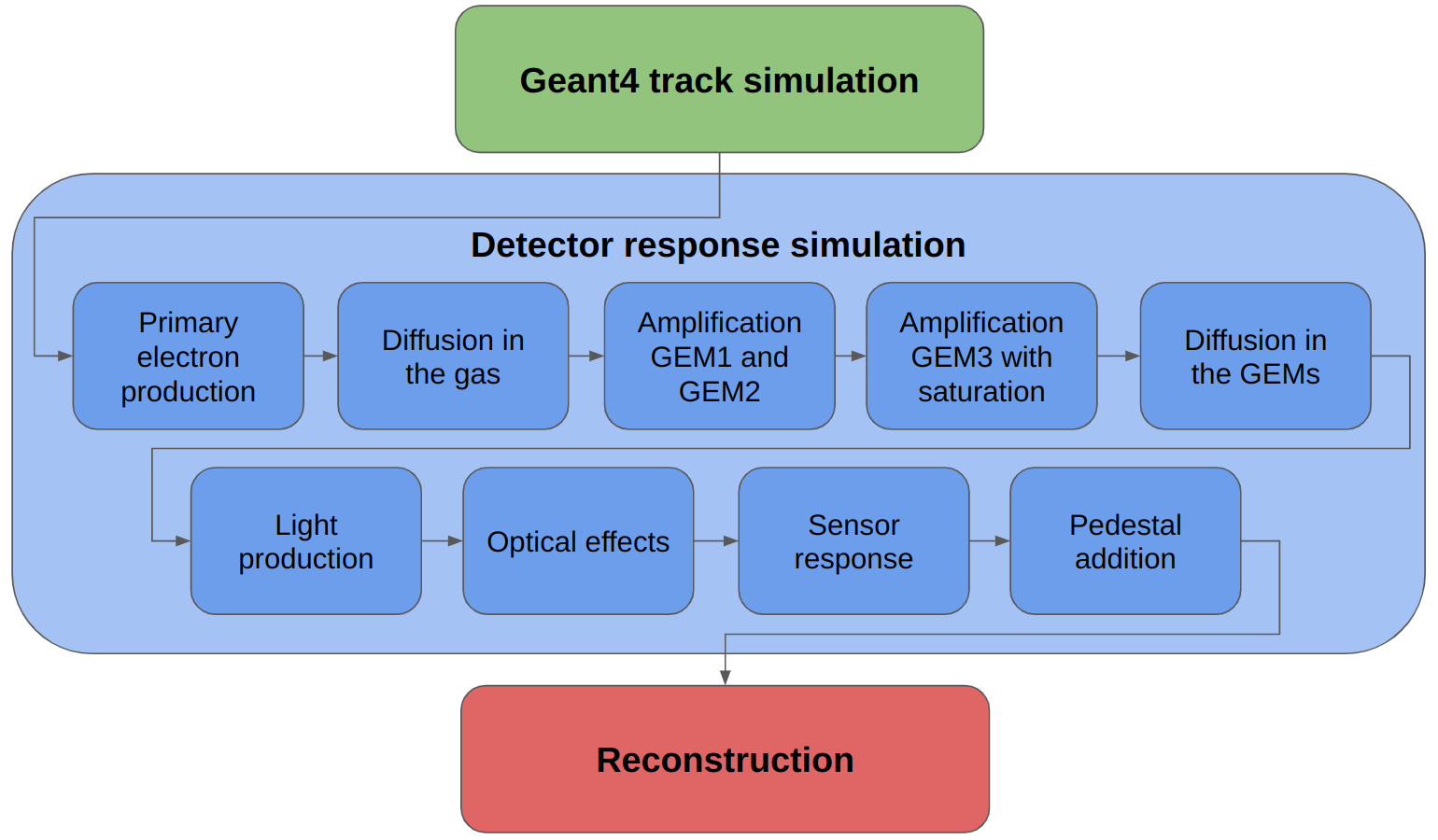}}
   \caption{Schematic representation of the simulation framework, as detailed in the text. 
   }
   \label{scheme_sim}
\end{figure}

The simulation proposed in this work is done through several steps that are schematically reported in Figure~\ref{scheme_sim}.

To simulate the detector response to ER interactions of the X-ray sources, we started modeling the energy deposits by using the \texttt{Geant4} toolkit \cite{geant} with the Standard EM Physics option4 model, selecting the most accurate physics models from Standard and Low Energy sets. The tracks were generated in a gas mixture of He and CF$_4$ with a 60:40 ratio. The total gas density was calculated as \SI{1.51}{\kilo\gram\per\cubic\metre} under typical operating conditions of \SI{970}{\milli\bar} to \SI{1000}{\milli\bar} pressure and \SI{295}{\kelvin} to \SI{300}{\kelvin} temperature. 
The simulation of the ERs included standard transportation, multiple Coulomb scattering, ionization, and bremsstrahlung. These processes collectively reproduced electron behavior in the CYGNO gas mixture. 

To reduce computational load, we avoid simulating the X-ray interactions and we inject directly the primary electrons with a kinetic energy equal to that of the X-ray line.
While the photoelectric effect has an angular dependence on both the energy and direction of the incident X-ray photon, for this analysis, the initial electron momentum directions were sampled isotropically. The primary interactions were initially simulated at the center of the sensitive volume, since at the level of the primary interaction all points in the TPC are equivalent. Spatial dependencies, such as gain non-uniformity, and diffusion effects, were subsequently implemented through appropriate translations and corrections in later stages of the simulation. This improves the computational efficiency in samples generation without loss of generality.

\begin{figure}[b]
 \centering
  {\includegraphics[width=100mm]{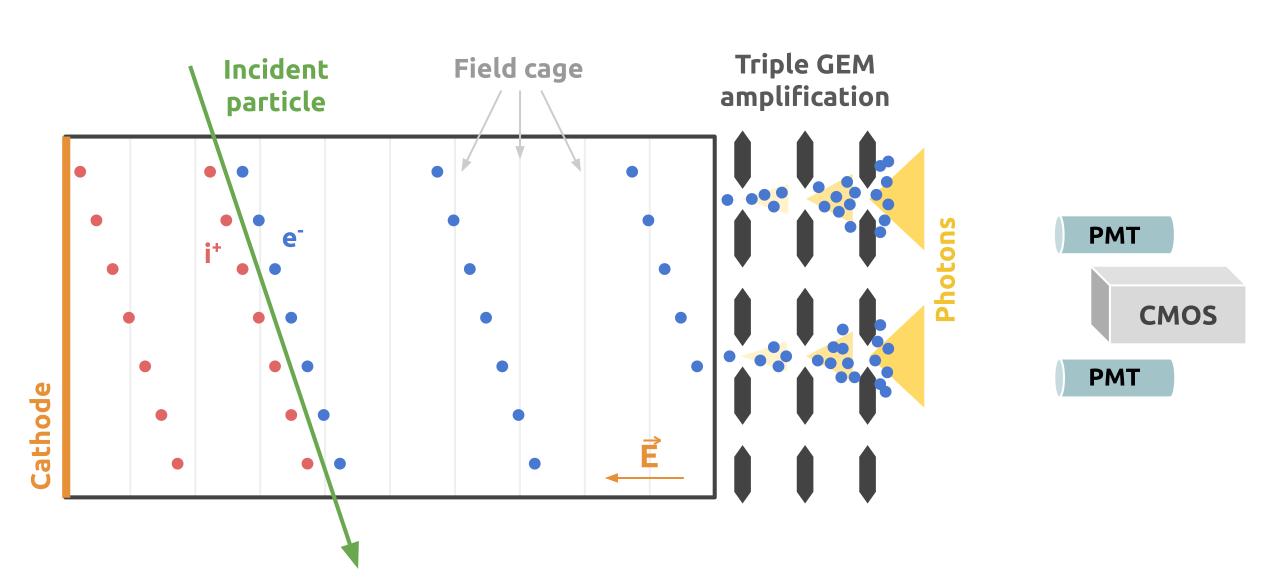}}
   \caption{Schematic representation of the TPC operation principle~\cite{MarquesThesis}. Following an ionizing event, the freed electrons drift under an electric field toward the amplification region, where they are multiplied and finally detected. In the CYGNO setup, this process takes place through a stack of GEMs, while the resulting signals are read by CMOS cameras and PMTs.
   }
   \label{TPC}
\end{figure}

The second part of the simulation aims to reproduce the detector response, as schematically shown in Figure~\ref{TPC}. 
An ionizing particle (green arrow) traverses the detector, ionizing the gas along its path and liberating primary electrons (blue dots). These ionization electrons drift along the electric field lines in the $z$-direction toward the amplification and readout plane ($x$-$y$ plane in our reference system), where a GEM stack amplifies the signal, producing photons in the multiplication process. sCMOS cameras and PMTs eventually read out these photons. 

Starting from the ER tracks simulated in \texttt{Geant4}, specifically from the information on the deposited energy along the track, our goal is to reproduce an image as it would appear from the sCMOS sensor output. All the relevant steps are detailed in the following of this section. The image is represented as a 2D histogram, where each cell contains the number of photons detected by the sensor at a given pixel, as in data. For our initial simulation studies, each simulated image contains only a single track. Although this is an approximation, it is justified by the low occupancy observed in real data. 
For instance, during the acquisitions with the X-ray sources (see Section \ref{data_samples}), the average event rate is approximately 250 Hz, corresponding to about 12–13 tracks per image. Considering that the image area is on the order of \(1000\, \text{cm}^2\), and that each 
reconstructed cluster corresponding to one low energy ER event
typically covers an area of approximately \(1\, \text{mm}^2\), the total area occupied is very small compared to the full image. Consequently, the probability of cluster overlap is low. 
Nonetheless, to further enhance the realism and completeness of the simulation study, incorporating pile-up effects would be beneficial, as these can influence the reconstruction process and introduce additional systematic uncertainties.
Finally, the simulated images are reconstructed using the same analysis code as in the data analysis to extract useful features \cite{dbscan}.

\subsection{Primary electrons production and absorption}
\label{prod_att}
For each $hit$ in the \texttt{Geant4} simulation, defined by an energy deposit $\Delta E$ at a specific position ($x$,~$y$,~$z$) along the track, the average number of ionization electrons produced, $\bar{N_e}$, is proportional to the energy deposit and is given by $\bar{N_e} = \Delta E/W_i$, where $W_i$ is the W-value of the gas, and represents the mean energy
required to create an electron-ion pair. A value of \SI{35}{\eV\per pair} was assumed, as obtained by taking the weighted mean \cite{Wvalue_mixt} of $W_i^{\text{He}} = \SI{41.3}{\eV\per pair}$ \cite{WvalueHe}, and  $W_i^{\text{CF}_4} = \SI{34}{\eV\per pair}$ \cite{Wvalue, Wvalue_thesis}.
The actual number of ionization electrons, $N^{pr}_e$, is then obtained by random sampling from a Poisson distribution with an expected value of $\bar{N_e}$, for simplicity.

During the drift towards the GEMs, there is a significant probability that electrons are captured by interactions with molecules in the gas volume. 
In the simulation, this effect is implemented by introducing an attachment process characterized by the attenuation length $\lambda$. Consequently, the number of electrons reaching the amplification plane is expressed as

\begin{equation}
    N_e = N_e^{pr} \cdot e^{-\frac{z}{\lambda}},
    \label{eq:attachment}
\end{equation}
where $z$ denotes the distance from the hit position to the amplification plane.

For highly pure gas mixtures, electron absorption rates as low as $3\%$/m are reported ~\cite{Blum2008}.
However, gas contamination due to imperfect gas tightness can increase this effect. In particular, even a small concentration of O\textsubscript{2}, that is highly electronegative~\cite{SnowdenIfft2013}, can substantially enhance electron attachment, reducing $\lambda$ to the order of a few meters. 
In our case, a $\lambda$ of the order of 1 meter is assumed and it is eventually optimized to achieve agreement with experimental data, as detailed in Section~\ref{sec5}.

\subsection{Electrons amplification and GEM gain}
\label{sec:ampl}
The ionization electrons, denoted by $N_e$, that reach the GEMs are subsequently amplified. Within the GEM holes, the electric field is extremely intense, reaching values up to $\sim 50\,\mathrm{kV/cm}$ \cite{light_yield_enhance}.

Such a strong field is sufficient to accelerate electrons, imparting them with enough energy to ionize the gas. This ionization process releases additional free electrons, which are then accelerated by the electric field to further ionize the gas via a Townsend avalanche mechanism. Considering that $n_{\mathrm{in}}$ electrons enter a GEM hole and $n_{\mathrm{out}}$ electrons are extracted, the gain $G$ of the GEM is defined as 
\[
G = \frac{n_{\mathrm{out}}}{n_{\mathrm{in}}}.
\]
The single GEM gain, as reported in \cite{lemon5}, can be parameterized, for the electric field configuration of the TPC, as a function of the GEM voltage as
\begin{equation}
G = 0.034 \cdot e^{0.021 \cdot \Delta V\mathrm{(V)}},
\label{eq:5.4}
\end{equation}
which encapsulates the exponential dependence on the applied voltage. This parametrization also includes the collection efficiency (i.e., the probability that an electron reaching a GEM foil enters a GEM channel).

Throughout the amplification process, electron absorption and imperfections in GEM manufacturing can affect the effective field and, as a consequence, the GEM gain, resulting in fluctuations. As reported in \cite{thesis}, the effective gain of the GEM follows an exponential distribution with a mean value equal to the average gain reported in Eq.\,\ref{eq:5.4}. These gain fluctuations are considered significant only in the first amplification stage, where the number of primary electrons is relatively low. After this initial stage, the electron count increases by $\sim$300 (which corresponds to a single GEM gain at $440\,\mathrm{V}$). Due to the large number of electrons, applying the same amplification process with gain fluctuations would have a negligible effect, as the fluctuations tend to average out. Therefore, in the subsequent amplification stages, these fluctuations are less critical and are not simulated in detail, mainly to reduce computational complexity.

Not all electrons generated during the amplification process successfully exit the GEM hole; some may be absorbed by the GEM surface. This efficiency, referred to as the extraction efficiency $\epsilon_{\mathrm{extr}}$, is dependent on the GEM voltage and, for the electric field configuration of the TPC, can be parameterized as follows \cite{GEMeffi1,GEMeffi2}:
\begin{equation}
\epsilon_{\mathrm{extr}} = 0.87 \cdot e^{-0.002 \cdot \Delta V\mathrm{(V)}},
\label{eq:5.5}
\end{equation}
showing a slight decrease in efficiency as the voltage increases.

In the simulation, for each point along the track containing $N_e$ primary ionization electrons, the number of electrons produced by the amplification of the first GEM, $N_e^{G_1}$, is calculated as:
\begin{equation}
N_e^{G_1} = \sum_{k=1}^{N_e} \left( G_k^{G_1} \cdot \epsilon_{\mathrm{extr}}^{G_1} \right),
\label{eq:5.6}
\end{equation}
where $G_k^{G_1}$ is the actual gain of the GEM for each electron, sampled from an exponential distribution with a mean value of $G^{G_1}$ (as given in Eq.\,\ref{eq:5.4}), and $\epsilon_{\mathrm{extr}}^{G_1}$ is the extraction efficiency, which represents the probability for an electron to exit the GEM channel. Since gain fluctuations are not considered beyond the first stage, the number of electrons at the second amplification stage is given by
\begin{equation}
N_e^{G_2} = N_e^{G_1} \cdot \left( G^{G_2} \cdot \epsilon_{\mathrm{extr}}^{G_2} \right).
\label{eq:5.7}
\end{equation}

The third and final amplification stage, which is subject to saturation effects, will be discussed in detail in the next section.

\subsection{Electrons diffusion}
During the drift in the gas, primary electrons experience diffusion. The transverse diffusion, $\sigma_{x,y}$, is modeled as a combination of diffusion due to electrons traveling through the gas volume, $\sigma_T$, and a fixed contribution associated with the GEM amplification, $\sigma_{T0}$. Similarly, the longitudinal diffusion, $\sigma_z$, is described following \cite{Blum2008}:
\begin{equation}
\sigma_{x,y} = \sqrt{\sigma_T^2 \cdot z + \sigma_{T0}^2}
\quad\quad
\sigma_z = \sqrt{\sigma_L^2 \cdot z + \sigma_{L0}^2},
\label{eq:5.8}
\end{equation}
with the diffusion coefficients $\sigma_T$ and $\sigma_L$ determined using Garfield++ simulations, varying with the applied drift field as reported in \cite{lemon5}, while $\sigma_{T0}$ and $\sigma_{L0}$ were measured from data in~\cite{saturation_davideP} and ~\cite{Folcarelli_thesis} respectively. 
The actual values of these parameters used in the simulation were optimized to match the data (see Section~\ref{sec:optimization}) starting from initial values from the references above (see Table~\ref{tab:params}).
In the simulation, the diffusion effect is implemented by smearing the $x$, $y$, and $z$ coordinates of each of the $N_e^{G_2}$ electrons around every interaction point along the track. The smearing is performed using a Gaussian function with a standard deviation of $\sigma_{x,y}$ for the $x$-$y$ coordinates and $\sigma_z$ for the $z$ coordinate. The resulting electron positions are stored in a three-dimensional histogram with voxel sizes $V_x$, $V_y$, and $V_z$. The voxel dimensions in $x$ and $y$ are chosen to match the area seen by each pixel of the optical sensor, while $V_z$ is selected to be on the order of the GEM thickness and is further optimized as described in Section~\ref{sec5}.

In a real TPC, diffusion in the drift occurs before entering the GEMs, and each amplification stage can introduce further spreading. In the simulation, however, the diffusion is implemented after the first two amplification stages in an uncorrelated way. Since the diffusion parameters were estimated from data, where the total diffusion is measured downstream of the amplification, this simplified ordering reproduces the observed diffusion behavior for low-energy X-ray events. 

\subsection{GEM gain saturation}\label{sec:saturation}
Studies have shown that when the electron cloud surpasses a certain density, the usual exponential dependence of GEM gain on voltage no longer holds \cite{limeLNF, saturation_davideP}. In particular, at the third GEM stage, the charge density can grow to the point where the electric charge partially shields the electric field inside the hole, causing a reduction in field strength.

To capture this behavior, a charge-density-dependent modification of the GEM gain has been introduced following the model detailed in \cite{saturation_davideP}. 
According to this model, we can write the total number of electrons exiting one GEM channel as 
\begin{equation}
n_{\mathrm{out}} = 
\frac{n_{\mathrm{in}}\,e^{\tilde{\alpha}\,(\Delta V-V_{0})}}
     {1 + \beta\,n_{\mathrm{in}}\bigl(e^{\tilde{\alpha}\,(\Delta V-V_{0})} - 1\bigr)},
\label{eq:5.11}
\end{equation}
where \(n_{\mathrm{in}}\) is the number of electrons entering the hole, $\tilde{\alpha}$ and $V_{0}$ are constants used for a linear parametrization of the Townsend coefficient 
and \(\Delta V\) is the voltage drop applied at the GEM plane.

In Eq.\,\ref{eq:5.11}, \(\beta\) determines the critical electron population at which amplification is lost, and was found to be of order 10$^{-5}$~\cite{saturation_davideP}. For \(n_{\mathrm{in}} = 1/\beta\), no net amplification is achieved. Letting \(g = e^{\tilde{\alpha}\,(\Delta V-V_{0})}\) be the unsaturated gain, Eq.\,\ref{eq:5.11} can be rewritten as
\begin{equation}
G(n_{\mathrm{in}}) = \frac{n_{\mathrm{out}}}{n_{\mathrm{in}}}
= A \cdot g 
\;\biggl[\,1 + \beta\,n_{\mathrm{in}}\,(g - 1)\biggr]^{-1},
\label{eq:5.12}
\end{equation}
where \(A\) is an overall normalization factor of order unity.

This adjustment is applied exclusively in the third amplification stage, where the electron population can reach approximately $10^5$. Consequently, each voxel of the 3D electron cloud is scaled by \(G_{G3}\bigl(n_{\mathrm{in}}\bigr)\). To obtain the total number of electrons exiting the third GEM, \(N_e^{\mathrm{out}}\), for each point in the \(x\)-\(y\) plane, the 3D cloud is projected along the \(z\)-axis onto the \(x\)-\(y\) plane.

\subsection{Light production and sCMOS sensor response}\label{sec:optics}
Due to the scintillating properties of the He:CF$_4$ gas mixture, the avalanche process is inherently accompanied by the emission of secondary photons alongside the avalanche electrons. The number of photons emitted is proportional to the number of electrons involved in the avalanche process.
Because the electron population after the first two amplification stages is significantly smaller than that produced after the third GEM (the gain achieved one stage earlier is about 100 times smaller, making the light coming from previous stages negligible), and since the third GEM effectively shields the optical sensors from the light generated in the preceding stages (the optical transparency of a GEM foil is evaluated to be $\sim$10\%), it is reasonable to assume that the photons reaching the sensors predominantly originate from the electrons emerging from the third GEM, \(N_e^{\mathrm{out}}\). In the simulation code, these electrons are eventually converted into scintillation photons, which are finally detected by the camera as ADC counts. The mean photon yield, \(\overline{N_\gamma}\), is
\begin{equation}
\overline{N_\gamma} = \mathrm{LY}\;\cdot\; N_e^{\mathrm{out}},
\label{eq:5.13}
\end{equation}
where \(\mathrm{LY} = 0.07\,\mathrm{ph/e^-}\) is the light yield of the gas mixture at atmospheric pressure \cite{Marafini2015, Mazzitelli2017, gem2}. Since photon production arises from gas molecule dissociation, it is treated as a Poisson process. Consequently, the actual number of photons, \(N_\gamma^{\mathrm{tot}}\), is drawn from a Poisson distribution with mean \(\overline{N_\gamma}\).
As detailed in Section~\ref{sec:ampl}, each bin in the 2D projection corresponds to the same physical area as one camera pixel, establishing a direct 1:1 correspondence. 

The optical system focuses these photons onto the sensor, but only a fraction reach the lens due to the limited solid angle. This fraction is represented by \(\Omega\), giving \(N_\gamma^{\mathrm{pix}} = N_\gamma^{\mathrm{tot}} \cdot \Omega\). The solid angle \(\Omega\) is defined as
\[
\Omega = \frac{1}{\Bigl(4\,N\bigl(\tfrac{1}{I} + 1\bigr)\Bigr)^2},
\]
where $N=0.95$ is the focal number and $I=0.042$ is the magnification \cite{CYGNOpaper}, resulting in ${\Omega=1.13\times 10^{-4}}$.

In order to reproduce real images more accurately, the \emph{vignetting} effect must also be included. This effect manifests as a decrease in brightness near the image edges compared to the center, commonly approximated by \(\cos^4(\phi)\), where \(\phi\) is the angle between the optical axis and the incoming rays \cite{goossens2019vignetted}. In a single-lens setup, natural vignetting alone can reduce the intensity at the borders by about 20\%. Moreover, in CYGNO the gain non-uniformities across the GEM surface introduce additional spatial variations in light production. Hence, each pixel’s photon count is multiplied by a response map that accounts for both the lens vignetting and the GEM response variations.

\begin{figure}[b]
 \centering
  \includegraphics[width=100mm]{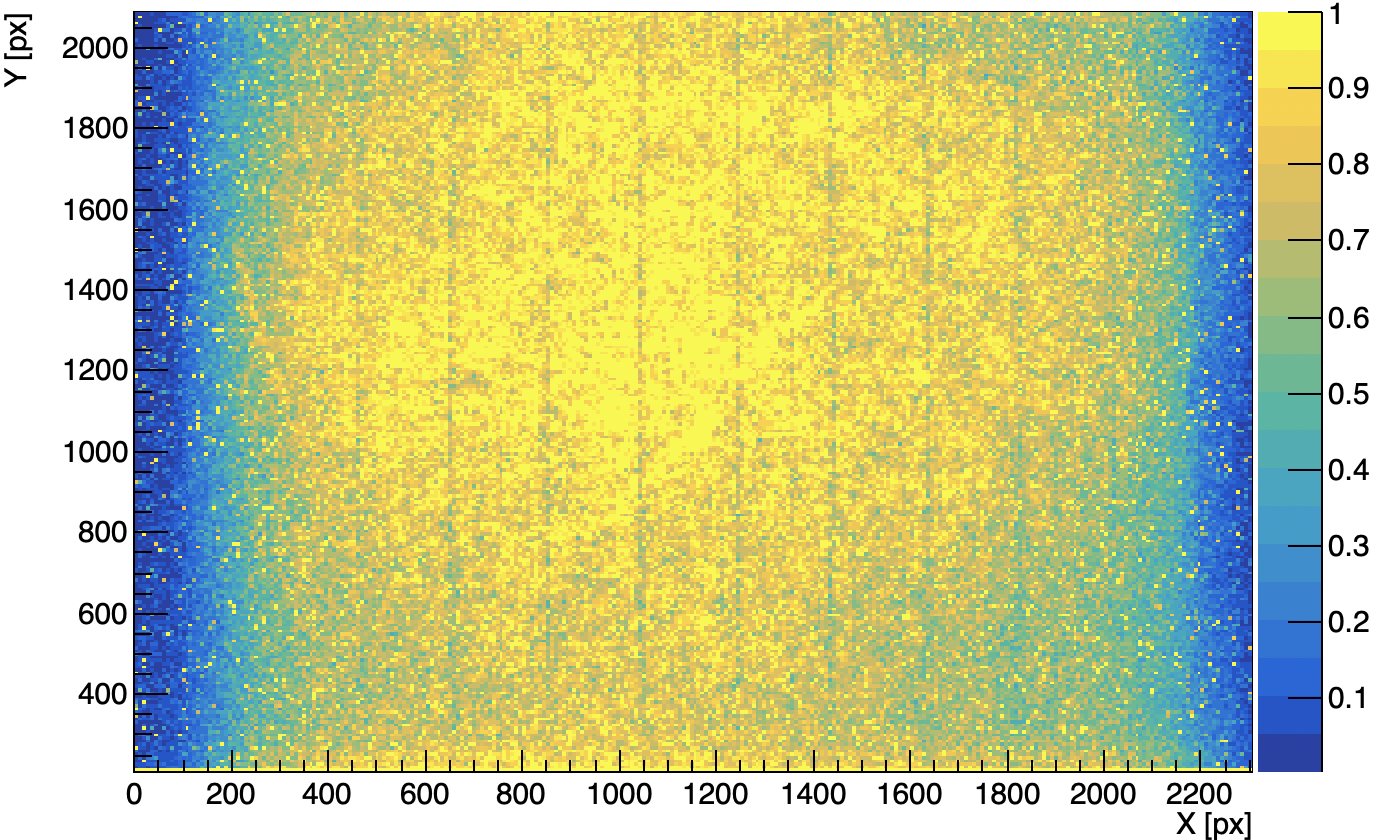}
   \caption{Example of the 2D response map obtained as described in the text. The color map represent the response of each pixel normalized to the light of the most intense one. Each pixel images an area of $150\times150~\mu \text{m}^2$ on the GEM plane. This map is used to simulate effects due to the gas-gain non-uniformities, the vignetting, and drift-field non uniformity.}
   \label{cosmic_map}
\end{figure}

To construct this map, a thousand images with a 2~s exposure time were collected and summed with the LIME detector recording events originating from internal and external radioactivity including cosmic rays. Each image was previously corrected, subtracting \textit{pedestal images} to remove fixed-pattern noise and other systematic effects unrelated to the physical signal.     
These pedestal images are acquired in dedicated runs where no external light is present, and the counts on the camera pixels originate solely from the intrinsic electronic noise of the optical sensor (including both readout noise and dark current). In these runs, the GEM voltage was set to 200 V a value low enough not to generate signals in the gas but close enough to those used during data acquisition, allowing for an accurate characterization of the sensor baseline. Conversely, the drift field was kept equal to that used during regular data taking. As a final step, the content of each pixel in the 2D map was normalized to the light of the most intense pixel.
The resulting map, shown in Figure~\ref{cosmic_map}, inherently includes both lens vignetting and potential GEM response non-uniformities, enabling more precise comparisons between data and simulation. 
The vertical lines appearing in the map, with a distance of approximately 200 pixels, are due to actual segmentation in the GEM lower electrode introduced to increase the stability of the detector (as described in Section~\ref{geant4_sec}).  

An alternative first-principles method was tested to evaluate the reliability of the map used to reproduce gain non-uniformities. Previous studies \cite{gain_disuniformity} have found that, for a 3-GEM stack of $10 \times 10\,\text{cm}^2$, the gain variation is around 8.8\% in $1.5 \times 1.5\,\text{cm}^2$ squares. As a cross-check, a gain fluctuation was applied per-track in the simulation without applying the map. The value of this fluctuation was optimized by comparing with the energy resolution as obtained from LIME data. The optimal value was found to be a 15\% gain fluctuation, which is consistent with the measurements in \cite{gain_disuniformity}. 
The consistent results between applying per-track gain fluctuations or using the 2D map supports the reliability of the latter method which is the one actually employed in our simulation framework.

\begin{figure}[ht]
    \centering
    \includegraphics[width=0.45\linewidth]{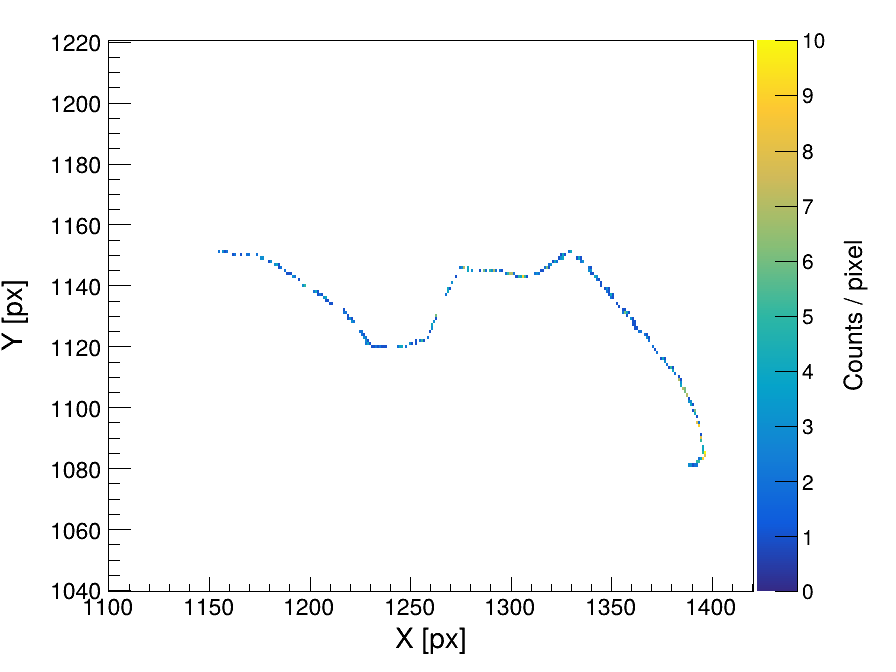}
    \includegraphics[width=0.45\linewidth]{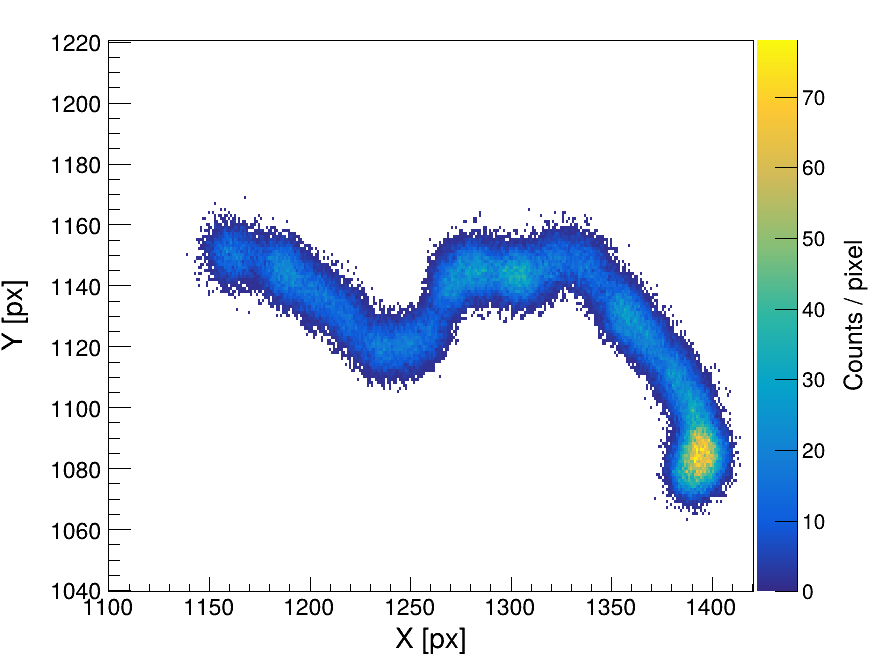}
    \includegraphics[width=0.45\linewidth]{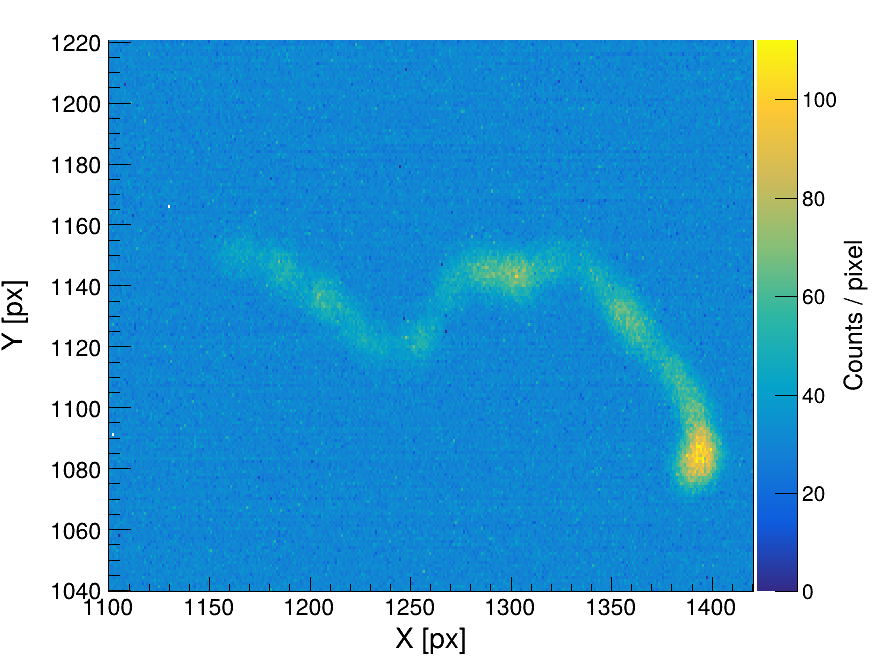}
    \caption{
        Detail of a picture from a \SI{60}{\keV} ER simulation: 
        (upper left) not considering diffusion effects in the gas and in the GEM stack; 
        (upper right) considering diffusion effects; 
        (lower) including sensor electronic noise in the final result.
    }
    \label{ER_pict_multi}
\end{figure}

In the final stage of the simulation, the sensor response is modeled. The camera converts the number of incident photons per pixel into ADC counts via a proportionality factor, \(C_{\mathrm{conv}} = 4\), provided by the manufacturer. The track image is then obtained as 
\[
\mathrm{ADC} = C_{\mathrm{conv}} \cdot N_\gamma^{\mathrm{pix}}.
\]
Next, the digitized track is superimposed onto a real sCMOS pedestal to create the final simulated image. Figure~\ref{ER_pict_multi} (bottom) shows an example of an electron recoil track produced in GEANT4, digitized, and with pedestal addition.

\newpage

\section{The LIME data samples analysis}
\subsection{Data samples}
\label{data_samples}
In a data taking campaign at LNF, data were collected using various X-ray lines to test the performance, linearity, and energy resolution of LIME, as well as to validate the simulation \cite{limeLNF}. A multi-target source was exploited, based on \(\alpha\) particles from \({}^{241}\)Am impinging on different materials (Cu, Rb, Mo, Ag, Ba, Tb) providing X-rays in the 8--50\,keV energy range. The \(\alpha\) particles ionize the inner shells of these materials, and the electrons refilling the vacancies produce the characteristic \(K_{\alpha}\) and \(K_{\beta}\) lines. The \(K_{\beta}\) lines have an intensity that is about 20\% of the corresponding \(K_{\alpha}\) lines. To obtain X-rays at about 5.9\,keV, a \({}^{55}\)Fe source was used. For lower energies, the same \({}^{55}\)Fe source was employed to excite a gypsum (Ca) target, yielding characteristic lines around 3.7\,keV. All these X-ray lines are summarized in Table~\ref{tab:single}, with each source placed at a distance of 25\,cm from the GEM stack, except for the \({}^{55}\)Fe source, which was also operated at 5\,cm, 35\,cm, and 45\,cm in additional runs to study the detector response at different drift distances.

In the case of the \({}^{55}\)Fe source, the active deposit is enclosed within a cylindrical lead container of 4.7 cm depth that fully blocks X-rays except those directed through an aperture toward the sensitive gas volume with a width of 1\,cm and 0.2\,cm along the \(x\) and \(z\) directions, respectively. 
Given the container depth and the 18 cm distance between the source support and the sensitive gas, geometrical considerations indicate that the resulting electron recoils (ER) are expected to have a maximum lateral spread along the \(x\) direction ranging from 4 cm (in the upper volume, closer to the source) to 11 cm (in the lower volume) from the central source position along the drift volume. Due to X-ray absorption in the gas, these events are predominantly concentrated in the upper part of the detector, closer to the source, where the spread is smaller. These considerations are consistent with the observed spread in the \(x\) direction, for which the \({}^{55}\)Fe track distribution has a standard deviation of approximately 5 cm. Along the \(z\) direction, being only 0.2\,cm the source aperture, the expected spread ranges from 1\,cm  to 2\,cm in the upper or lower parts of the gas volume, respectively. 

\begin{table}[ht]
\centering
\begin{tabular}{l c c c}
\hline
\textbf{Source} 
& \(\mathbf{E_{K_\alpha} [keV]}\)
& \(\mathbf{E_{K_\beta} [keV]}\)
& \textbf{Distance from GEM [cm]} \\
\hline
Ca\(\,^*\) 
& 3.69 
& 4.01 
& 25 \\
Fe 
& 5.9 
& 6.5 
& 5--45 \\
Cu\(\,^\dagger\) 
& 8.04 
& 8.91 
& 25 \\
Rb\(\,^\dagger\) 
& 13.37 
& 14.97 
& 25 \\
Mo\(\,^\dagger\) 
& 17.44 
& 19.63 
& 25 \\
Ag\(\,^\dagger\) 
& 22.10 
& 24.99 
& 25 \\
Ba\(\,^\dagger\) 
& 32.06 
& 36.55 
& 25 \\
Tb\(\,^\dagger\) 
& 44.23 
& 50.65 
& 25 \\
\hline
\end{tabular}
\caption{
\label{tab:single}
Combined X-ray lines. 
\(\dagger\)~indicates multi-target sources (Cu, Rb, Mo, Ag, Ba, Tb). 
\(*\)~indicates the Ca line induced by \({}^{55}\)Fe fluorescence. 
The iron lines (Fe) at 5.9\,keV and 6.5\,keV come directly from the \({}^{55}\)Fe source.
Distances from the GEM are 25\,cm for most sources, while Fe was placed between 5\,cm and 45\,cm.
}
\end{table}

\subsection{Events reconstruction}
\label{reco}
Events were reconstructed using the official CYGNO reconstruction code\footnote{\url{https://github.com/CYGNUS-RD/reconstruction}}. After a pedestal subtraction of the image, a zero-suppression algorithm is applied to remove residual sensor noise, and an optical correction is performed to account for the effects discussed in Section~\ref{simulation}. 

The remaining fired pixels in the image are processed using a dedicated algorithm developed by the collaboration \cite{iddbscan}, based on an upgraded version of DBSCAN \cite{dbscan_alg}.
The algorithms identifies the pixels belonging to each the track and allows to compute the track shape variables used in the analysis:

\begin{itemize}
    \item \textbf{Light Integral:} The total light content of the track (\boldmath{$I_{SC}$}), given by the sum of all pixel counts, representing the overall energy deposit.
    \item \textbf{Transverse Gauss Sigma:} The standard deviation of a Gaussian fitted to the transverse profile of the track. This reflects the dispersion of energy deposition due to drift diffusion, GEM diffusion, and any curvature of the track.
    \item \textbf{Transverse Gauss Amplitude:} The maximum amplitude of the Gaussian fitted to the transverse profile of the track, indicating the brightness at the track’s center.
    \item \textbf{Transverse Gauss Mean:} The mean position of the Gaussian fitted to the transverse profile of the track, reflecting the position where most of the energy is released along the track.
    \item \textbf{Number of Hits:} The total count of pixels above the zero-suppression threshold. This quantity is directly related to the spread of the initial ionization, affected by diffusion in the gas and amplification stages.
    \item \textbf{Size:} The area (in pixel units) of the ellipse that encloses the track, reflecting its overall extent on the sensor plane.
    \item \textbf{Length:} The length of the major axis of the enclosing ellipse (in pixel units). Along with the track width, it provides a measurement of the track’s shape on the x-y plane.
    \item \textbf{Width:} The length of the minor axis of the enclosing ellipse (in pixel units). Together with track length, it characterizes the two principal dimensions of the projected cluster.
    \item \textbf{Slimness:} The ratio of track width to track length, offering a measure of the track’s roundness (a small value indicates an elongated shape).
    \item \textbf{Density:} The ratio of the total light contained in the track to its number of hits, indicating how concentrated the light/energy is within the cluster.
    \item \textbf{Average Specific Ionization:} The ratio of the light integral to the track length, quantifying how much light is produced per unit length of the cluster.
\end{itemize}

\subsection{Data analysis}\label{data_analysis}
In the analysis, an initial loose selection of the data was applied. Tracks with a length of less than \SI{400}{pixels} (equivalent to \SI{6}{\cm}) were selected to exclude longer tracks typically associated with ERs with an energy above $\sim$100~keV, originating from radioactive background or cosmic rays. To minimize the presence of noise caused by vignetting and other detector inhomogeneities, only tracks with a barycenter located within \SI{800}{pixels} (\SI{12}{\cm}) from the image center were included. To remove fake clusters - random groupings of neighboring pixels above the zero-suppression threshold that do not correspond to real signals - events with \( I_{SC} > \SI{1000}{counts} \) were selected, as discussed in \cite{limeLNF}.

Given the low interaction length of gamma rays from Ca, different selection criteria were used on the data: track length less than \SI{300}{pixels} (\SI{4.5}{\cm)}, and position selection within a 500$\times$200~pixels horizontal rectangle on the top side of the pictures.

\begin{figure}[ht]
 \centering
  {\includegraphics[width=100mm]{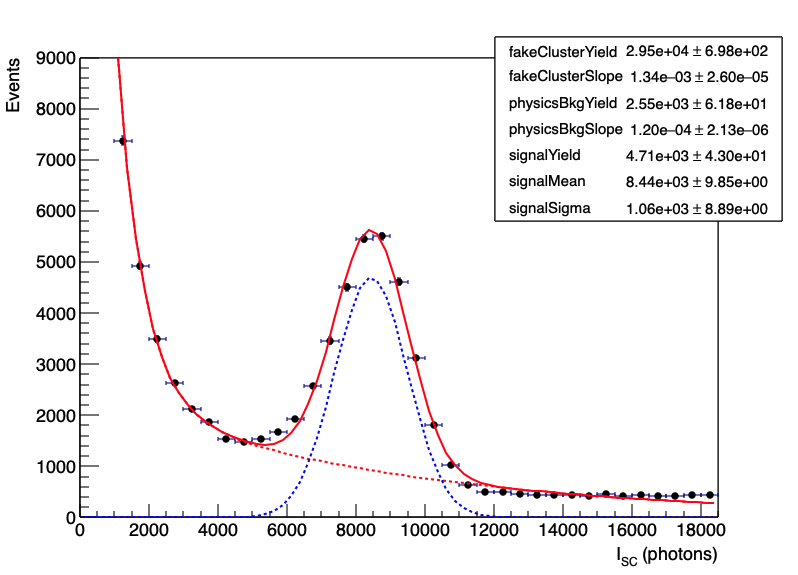}}
  \caption{Distribution of the light integral, $I_{SC}$, from data collected with the Cu source. The solid red line represents the overall fit, while the red dotted line corresponds to the background component, and the blue dotted line represents the signal component as described in the text. The value of the parameters as obtained from the fit are displayed in the figure.}
   \label{Cu_fit}
\end{figure}

The energy spectra of the different sources were fitted. A composite function, consisting of two exponential functions to model the background plus a Gaussian function to model the signal, has been used. The first exponential function is steeper and models the fake clusters reconstructed at lower energies, while the second models the physics background (cosmic rays and electrons from natural radioactivity). An example of the fit on the Cu data can be found in Figure~\ref{Cu_fit}.

\section{Simulation optimization and results}
\label{sec5}

The simulation described in Section~\ref{simulation} was optimized to reproduce the data sets acquired with the LIME detector at LNF, using only the $^{55}$Fe source placed at various distances from the GEM readout plane. Identical selection criteria were applied to both simulated and experimental data, as detailed in Section~\ref{data_analysis}, with an additional constraint: tracks with a slimness greater than 0.8 were excluded to preferentially select the compact, rounded clusters typical of X-ray interactions from the $^{55}$Fe source.

The optimization procedure focused on four reconstructed quantities (light integral, number of hits, transverse Gaussian amplitude, and transverse Gaussian sigma) as defined in Section~\ref{reco}. These quantities were found to be the most informative for the optimization process. For each source position along the $z$-axis and for each configuration of simulation parameters, the distributions of these four variables, obtained from approximately 200 reconstructed clusters, were fitted with a Gaussian function.

Although some distributions deviate from an ideal Gaussian shape, the aim was to capture the quasi-Gaussian core of each distribution and to use its mean value for comparing experimental data with simulation predictions. Further details on the comparison of track shape variable distributions between data and simulation are provided in Section~\ref{shape_comparison}.

The optimization strategy relies on minimizing a metric that quantifies the residuals between experimental data and simulation predictions. For each of the four key quantities, the squared difference between the mean simulation prediction and the corresponding experimental measurement was calculated at each source position. These squared differences were then summed over all source positions and all four quantities to yield a single figure of merit.

\subsection{Parameter optimization procedure}\label{sec:optimization}
The optimization followed a sequential approach where parameters were tuned one at a time while keeping others fixed. Starting from theoretically-derived initial values (listed in Table~\ref{tab:params}), each parameter was systematically varied within a specified range, with the best-fit value being determined before proceeding to the next parameter. The order of optimization is indicated by the \textit{Opt. Step} column in Table~\ref{tab:params}. Initially, the two transverse diffusion coefficients were optimized simultaneously, as illustrated in Figure~\ref{gas_opt} (left). Then, longitudinal diffusion coefficients were optimized in a similar way. Following the diffusion coefficients, the attenuation length $\lambda$, the saturation parameter $\beta$, and the normalization constant $A$ were optimized in sequence. Finally, the voxel dimension along the drift direction $\Delta z_{\text{vox}}$ was optimized, followed by the simultaneous optimization of the transverse voxel dimensions ($\Delta x_{\text{vox}}$ and $\Delta y_{\text{vox}}$), assuming equal dimensions in the transverse plane.

This approach acknowledges the possibility that different combinations of parameter values might yield similarly good agreement with experimental data. For example, adjusting the saturation parameter $\beta$, or the longitudinal diffusion parameters, can produce effects equivalent to modifying the voxel dimensions, illustrating the potential inter-dependencies between seemingly distinct parameters. A comprehensive exploration of such parameter inter-dependencies or alternative optimization sequences was beyond the scope of this work considering that the resulting set of parameters is found very close to the initial theoretically-derived values.

\begin{table}[ht]
\centering
\caption{Key parameters in the simulation. For each parameter, 
we list its initial value, the range used in the optimization scan, 
and the final best-fit value. The first column also indicates 
the step at which each parameter was progressively optimized.}
\label{tab:params}
\vspace{0.5 cm}
\begin{tabular}{c l c c c}
\hline
\textbf{Opt. Step} & \textbf{Parameter} & \textbf{Initial Value} & \textbf{Range} & \textbf{Best Value}\\
\hline
I   
   & $\sigma_T\,(\mu\mathrm{m}/\sqrt{\mathrm{cm}})$ & 120~\cite{lemon5} & 110--130 & 115 \\
   & $\sigma_L\,(\mu\mathrm{m}/\sqrt{\mathrm{cm}})$ & 100~\cite{lemon5} & 50--150  & 100 \\
& $\sigma_{T0}\,(\mu\mathrm{m})$                 & 450~\cite{saturation_davideP} & 300--550 & 350 \\
   & $\sigma_{L0}\,(\mu\mathrm{m})$                 & 250~\cite{Folcarelli_thesis} & 150--350 & 260 \\
\hline

II  & $\lambda\,(\mathrm{mm})$                        & 1000 & 800--1600 & 1400 \\
\hline

III & $\beta$                             & $10^{-5}$~\cite{saturation_davideP} & $0.5-5 \cdot 10^{-5}$   & $10^{-5} $   \\
\hline

IV  & $A$                                            & 1.4 & 1.2--1.6 & 1.5 \\
\hline

V   & $\Delta z_{\text{vox}}\,(\mathrm{mm})$         & 0.1& 0.05--0.25 & 0.1 \\
\hline

VI  & $\Delta x_{\text{vox}}\,(\mathrm{mm})$         & 0.15 & 0.1--0.3 & 0.15 \\
  & $\Delta y_{\text{vox}}\,(\mathrm{mm})$         & 0.15 & 0.1--0.3 & 0.15 \\
\hline
\end{tabular}
\end{table}

\subsection{Diffusion parameters}
\label{diffstudies}

\begin{figure}[ht]
    \centering
          {\includegraphics[width=0.497\linewidth]{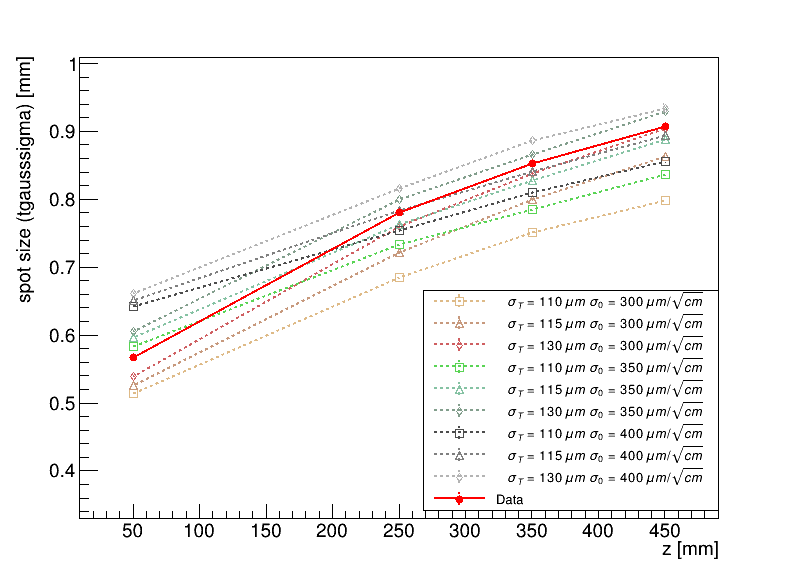}}
        {\includegraphics[width=0.497\linewidth]{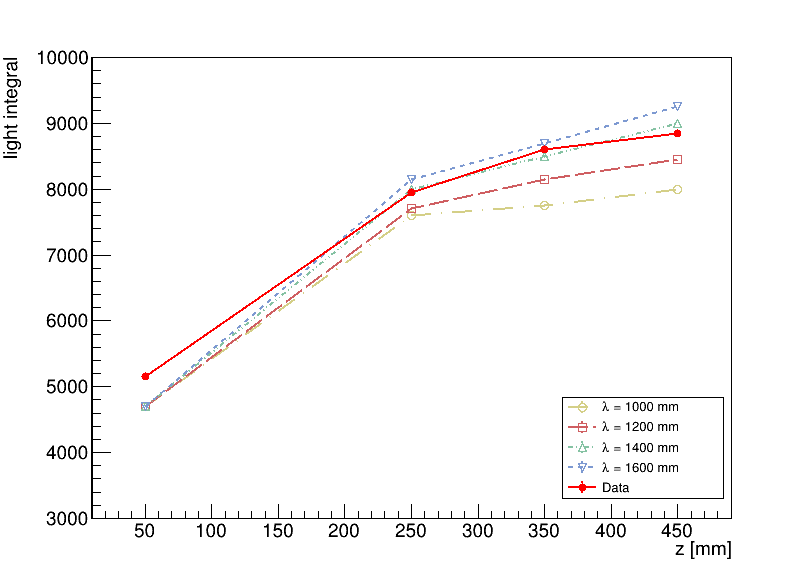}}
\caption{Optimization of the diffusion parameters and of the attenuation length: light integral as a function of the distance from the GEMs in the drift direction ($z$) for different values of the diffusion parameters (left) or different values of the attenuation length (right).}
\label{gas_opt}
\end{figure}

The diffusion parameters \(\sigma_{T0}, \sigma_{L0}, \sigma_{T}, \sigma_{L}\) were initially set around the Garfield-simulated values for a drift field of \SI{1}{\kilo\volt\per\centi\metre} at atmospheric pressure~\cite{lemon5}. These parameters were scanned in turn to find the best agreement with data, according to the sum of squared residuals described above. Figure~\ref{gas_opt} (left) shows how the cluster transverse size grows with \(z\), as expected from diffusive processes. At small values of \( z \), the transverse spread is dominated by \( \sigma_{T0} \), which accounts for the diffusion occurring within the GEM stack. As \( z \) increases, the contribution from \( \sigma_{T} \), associated with diffusion during the drift, becomes increasingly significant.

The resulting optimal diffusion values are comparable to Garfield predictions and are reported in Table \ref{tab:params}.

\subsection{Attenuation length}
In the pure gas mixture used in LIME at atmospheric pressure, the attenuation length---defined as the average distance a primary electron drifts before being absorbed by electronegative impurities---is expected to be of the order of \SI{1}{\metre} as discussed in Section~\ref{prod_att}. However, contamination from oxygen and other electronegative species can reduce this value. In the overground LIME setup exploited in the data taking considered in this work, direct measurements of contaminant concentrations were not available. Therefore, \(\lambda\) was scanned over a broad range, including values above \SI{1}{\metre} (up to and beyond \SI{1.4}{\metre}), to best match the experimental data. Figure~\ref{gas_opt} (right) illustrates the dependence of the light integral versus \(z\) on different \(\lambda\) values; larger deviations are observed at higher \(z\), where reabsorption effects are more pronounced. The best-fit value was found to be \(\lambda = \SI{1400}{\milli\metre}\).

\subsection{Saturation Parameters}
The saturation model introduced in Section~\ref{sec:saturation} treats each voxel as analogous to a GEM channel, with dimensions of approximately \SI{0.1}{\milli\metre} along \(z\) and \SI{0.15}{\milli\metre}~\(\times\)~\SI{0.15}{\milli\metre} in the transverse plane. As detailed in Section~\ref{sec:saturation}, two additional parameters—\(\beta\) and \(A\)—are required to describe the saturation effects. Both parameters were scanned within a range around their nominal values.

\begin{figure}[ht]
    \centering
        \includegraphics[width=0.497\linewidth]{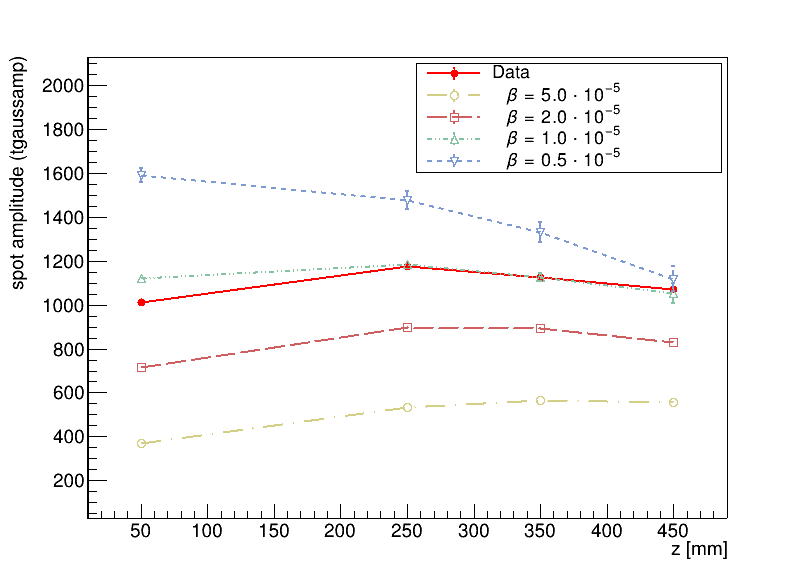}
        \includegraphics[width=0.497\linewidth]{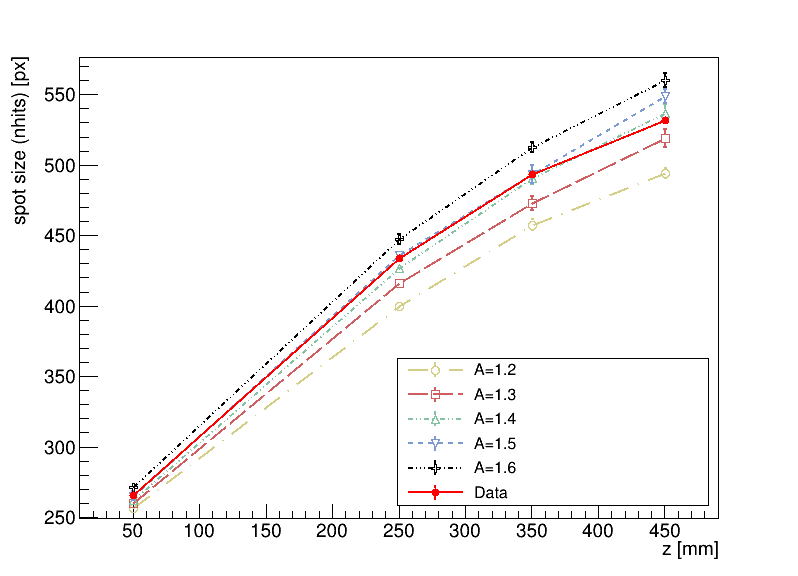}
\caption{Optimization study for the saturation parameters. Spot amplitude as a function of the distance $z$ for different values of the parameter $\beta$ (left) and spot size as a function of the distance $z$ for different values of the parameter A (right).}
\label{sat_opt}
\end{figure}

Figure~\ref{sat_opt} (left) shows how the transverse Gaussian amplitude depends on \(\beta\): smaller values of \(\beta\) enhance the saturation effect at low \(z\), where the electron cloud density is higher. Conversely, Figure~\ref{sat_opt} (right) illustrates that increasing \(A\) amplifies the spot amplitude more significantly at larger~\(z\).
The results of the optimization is reported in Table~\ref{tab:params}.

\section{Detector response and simulation comparison}

\subsection{Energy Resolution for Different Source Positions}
\begin{figure}[ht]
 \centering
  {\includegraphics[width=90mm]{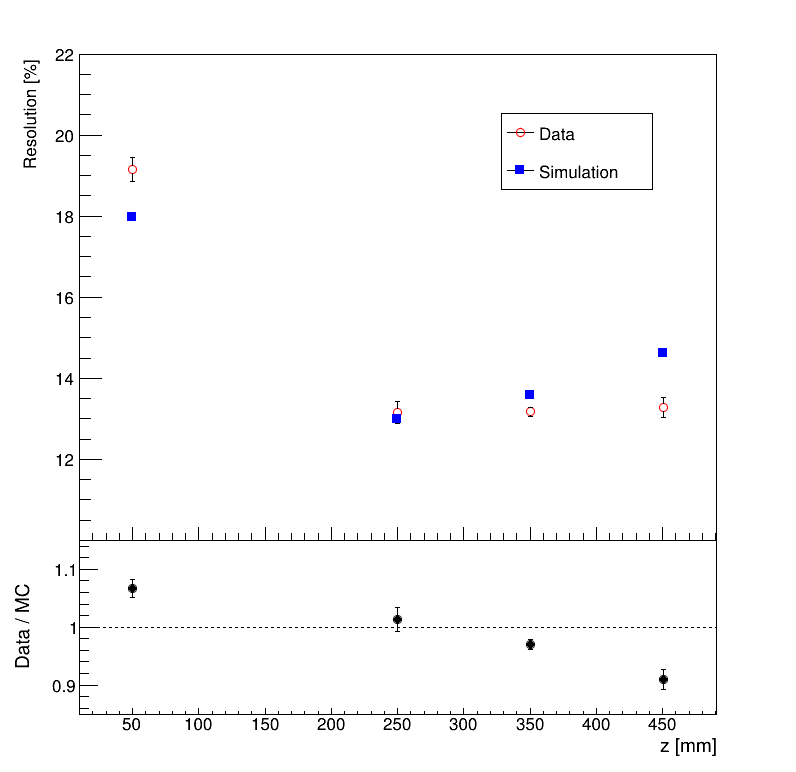}}
   \caption{Energy resolution of the 5.9~keV peak of the $^{55}$Fe X-rays as a function of the source position from the GEM as obtained in experimental data (red circles) and simulation (blue squares), highlighting their agreement across various positions.}
   \label{resolution_vs_z}
\end{figure}

The energy resolution is defined as the ratio of the standard  deviation, obtained from the Gaussian fit to the reconstructed energy peak, to the corresponding mean value.
The measured resolution for $^{55}$Fe at various source positions is shown in Figure~\ref{resolution_vs_z} and compared to simulation results obtained using the optimized saturation parameters, as described in Section~\ref{sec5}.

In the simulation, tracks were generated within the gas volume to accurately reproduce the spatial spread observed in data. They were uniformly distributed in the \(x\)–\(y\) plane within a radius of \SI{12}{\centi\metre} from the center of the detector, and with drift distances ranging from \SI{20}{\centi\metre} to \SI{30}{\centi\metre} along the \(z\)-axis.

The poor resolution observed at small \(z\) is attributed to stronger gain saturation effects in this region, where the charge density is higher. Additionally, the limited collimation of the source leads to a larger spread along the \(z\)-direction, which further degrades the resolution near the amplification plane.

\subsection{Detector energy response and resolution at different energies}\label{comparison} 
To compare data and simulation, high-statistic samples of electron recoil tracks at the energies of the different sources exploited in the data taking campaign were produced using the set of optimized simulation parameters. All tracks were reconstructed with the CYGNO reconstruction code using the same parameters used to reconstruct the data.

\begin{figure}[ht]
 \centering
  \includegraphics[width=0.49\linewidth]{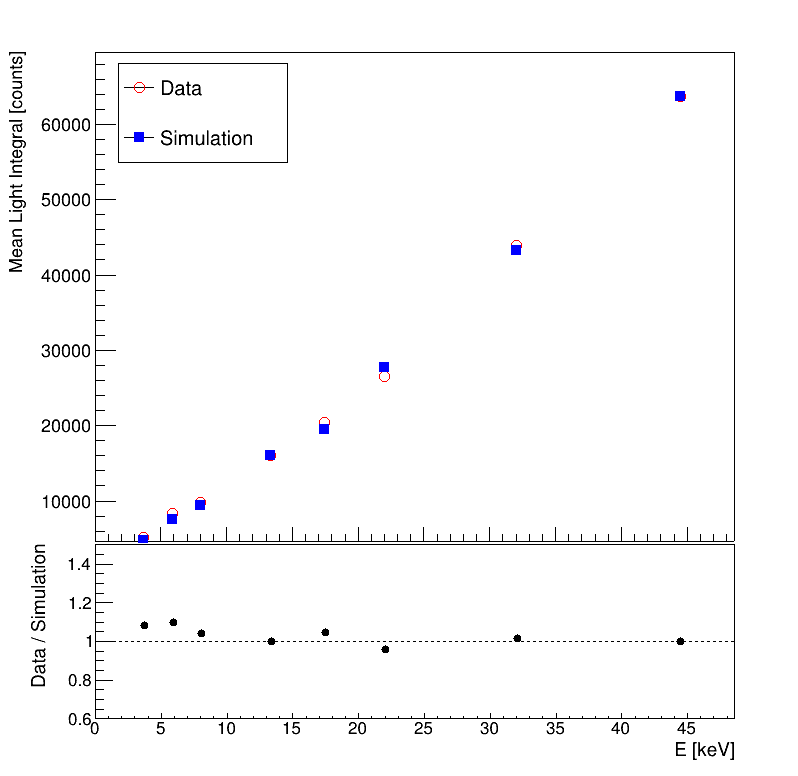}
  \includegraphics[width=0.49\linewidth]{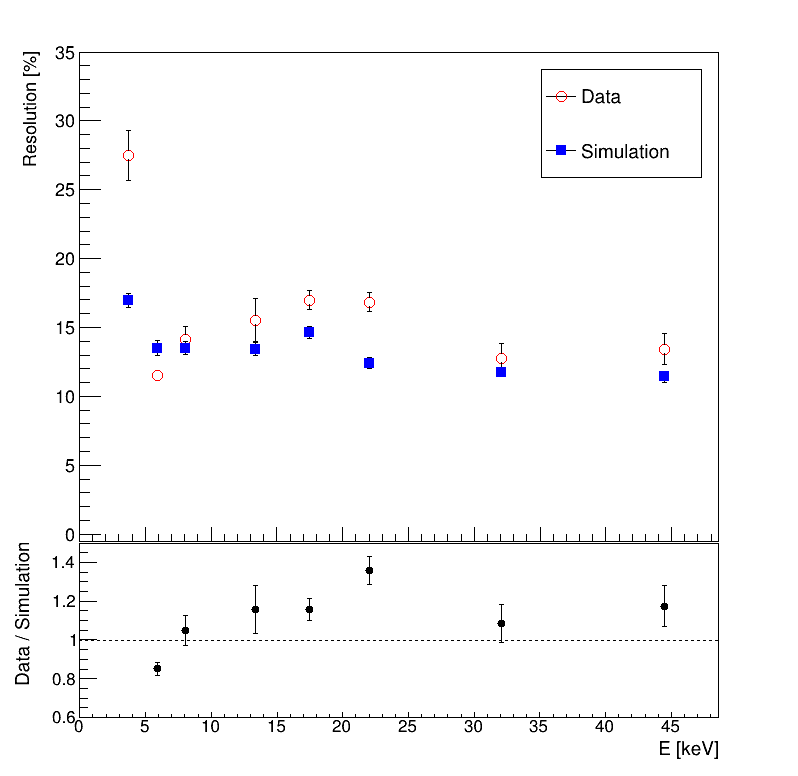}
  \caption{Data/simulation comparison of the detector response (left) and energy resolution (right) as a function of the energy. The mean light integral ($I_{SC}$) and resolution ($\sigma/I_{SC}$) are obtained for each energy value from a Gaussian fit to the light integral distributions, with error bars showing statistical uncertainties. Data points were collected using X-rays from $^{241}$Am on various targets (Cu, Rb, Mo, Ag, Ba, Tb) for 8, 13, 17, 22, 32, and 44 keV X-rays, $^{55}$Fe source for 5.9 keV X-rays, and $^{55}$Fe on Ca target for 3.7 keV X-rays. The energy of the dominant \(K_{\alpha}\) line peak is considered.}
   \label{data-mc_comp}
\end{figure}

The plots shown in Figure~\ref{data-mc_comp} prove a good agreement between data and simulation across a wide energy range. In particular, the mean reconstructed energy, Figure~\ref{data-mc_comp} (left), is  reproduced within a few percent, highlighting the robustness of the simulation model. The energy resolution, Figure~\ref{data-mc_comp} (right), is reproduced within 20-30\%. However, a noticeable discrepancy is observed in the resolution of the calcium. This deviation may point to additional sources of uncertainty, such as a significantly larger background or the difficulty of fitting a broad peak superimposed on a shoulder. Alternatively, it could indicate that certain systematic effects at low energies are not yet fully accounted for in the current simulation framework.

\subsection{Comparison of topological track shape variables}
\label{shape_comparison}
To study the accuracy in the reproduction of the tracks as they would appear in real sCMOS images, in addition to the light response and the energy resolution, a comparison between data and simulation of nine selected topological track variables has been carried out. 
However, unfolding the distribution of these shape variables from the data is challenging, mainly because of the presence of the background components.
To address this challenge and unfold the distributions for the signal component in the data, the $_s\mathcal{P}lot$ technique has been employed \cite{Pivk_2005}. 

When dealing with a data set of two uncorrelated variables, each comprising a signal and a background component, and the probability density function is known only for the first variable, $_s\mathcal{P}lot$ enables the extraction of the distribution of the second variable for the pure signal component in a statistical sense. This is achieved by conducting an unbinned likelihood fit on the first component. During fitting, each event is assigned two weights: $_s \mathcal{P}_s$ and $_s \mathcal{P}_b$, which are proportional to the probabilities of the event being signal and background, respectively. By building the distribution of the second variable that weights each event by $_s \mathcal{P}_s$, the distribution for the pure signal component can be obtained as demonstrated in \cite{Pivk_2005}. 
In each X-ray dataset, the light integral is used as the control variable since its p.d.f. is known. 
The variables chosen for the data-simulation comparison are the ones listed in Section~\ref{reco}. The pure signal is considered for the simulation since the data distributions are background subtracted with the sPlot technique.

\begin{figure}[ht]
 \centering
  {\includegraphics[width=150mm]{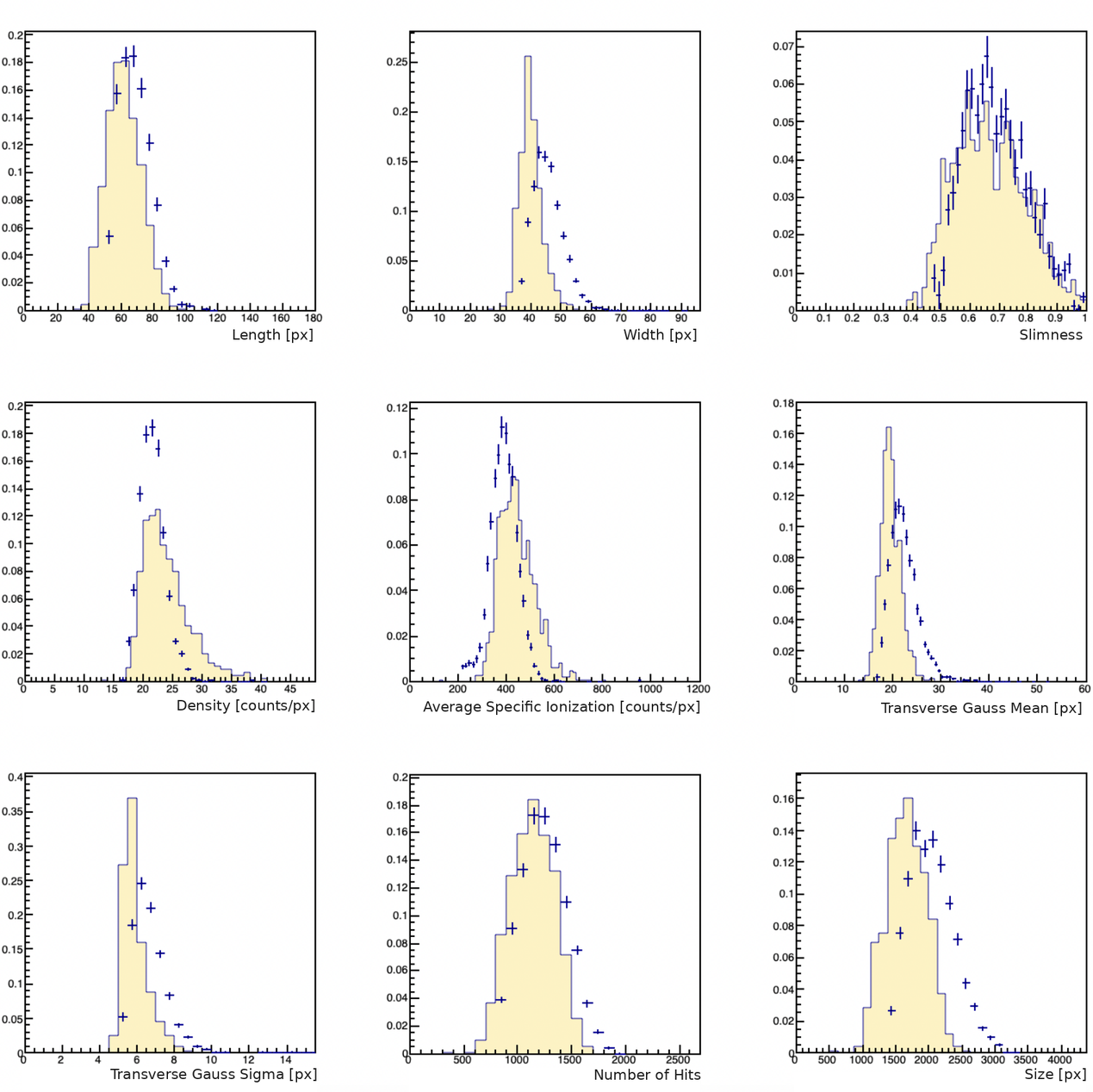}}
   \caption{Comparison of track shape variable distributions between data and simulation for ER tracks induced by \SI{24}{\keV} X-rays emitted from Ag. The histograms represent the data-unfolded signal distributions, constructed by weighting the quantity of each entry with $s P_s(y)$ (blue crosses), and the simulated tracks (yellow histograms).}
   \label{Ag_shapes}
\end{figure}

\begin{figure}[ht]
 \centering
  {\includegraphics[width=150mm]{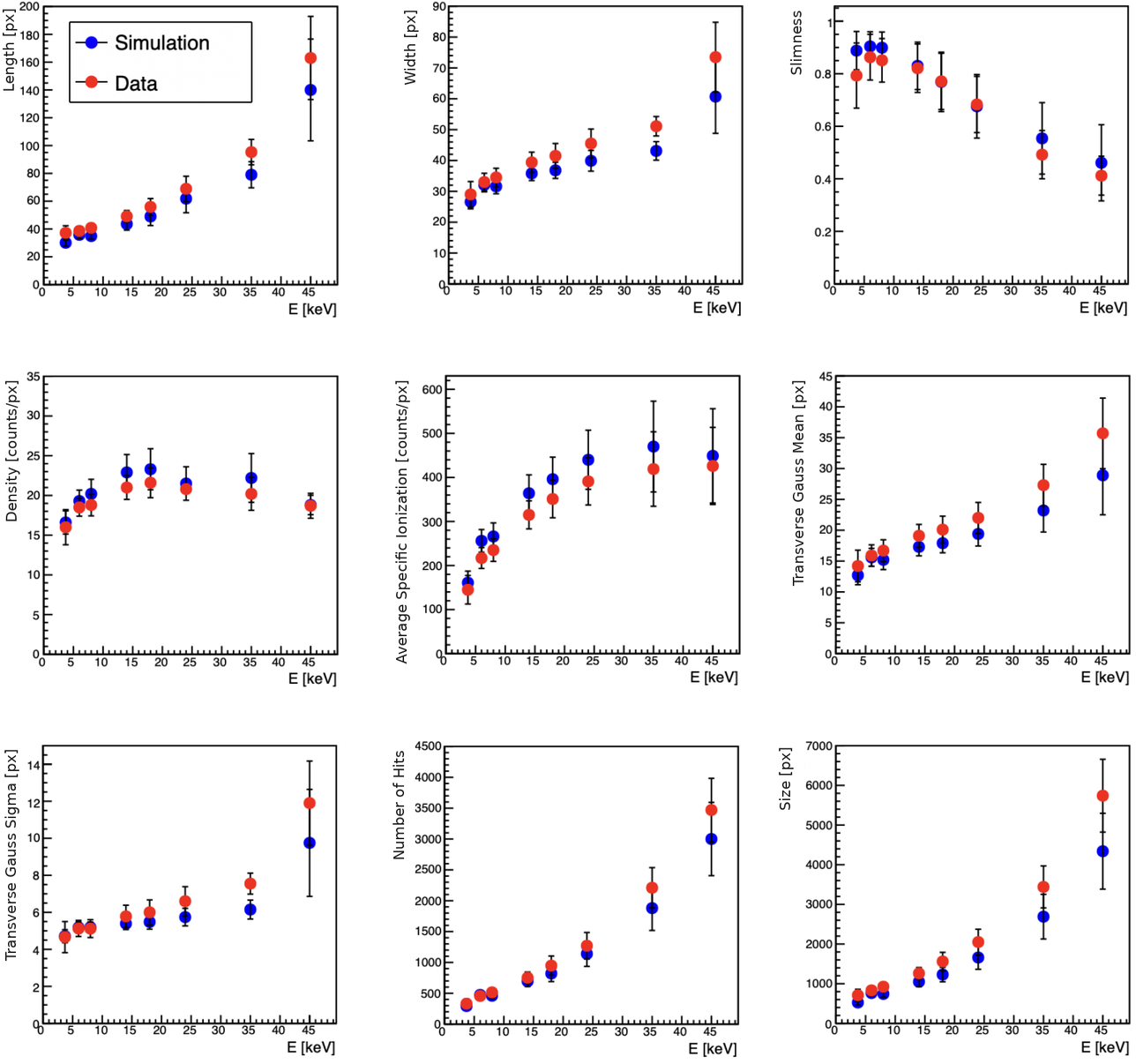}}
   \caption{Mean values of the track shape variables employed in the data-MC comparison, shown as a function of energy for each variable considered in the analysis. Blue markers indicate the simulated data, whereas red markers represent experimental data.}
   \label{shapes_vs_energy}
\end{figure}

Data-MC comparison is then performed between the pure signal distribution of these variables extracted from the data and the distributions of the simulated tracks. The normalized distributions are shown for the Ag data set as an example in Figure~$\ref{Ag_shapes}$. \\
Despite observing some systematic differences in distributions between data and MC, it is noteworthy to recognize the overall achievement given the complexity of the simulation, which aims to reproduce the entire electron recoil sCMOS image. The track length is well reproduced, as it is primarily governed by the underlying physics. For other variables, such as the width and size of the clusters, where the detector response (e.g., spatial resolution, diffusion) plays a significant role, the overall shape is well captured, but discrepancies in the average values suggest that further fine-tuning is needed, possibly in the noise modeling. Notably, the density---a critical variable for distinguishing ERs from NRs---shows a well-reproduced peak, though the shape could be improved. Similarly, $dE/dx$ demonstrates good agreement, which is crucial for modeling the track energy loss versus length and is valuable for head-tail discrimination. 
The distributions from data and simulation were also fitted with a Landau or a Gaussian depending on the distribution, to extract their central values and associated uncertainties. For each track shape variable, the resulting mean values and corresponding errors (taken as the standard deviation from the fit) were plotted as a function of energy for both data and Monte Carlo. Figure~\ref{shapes_vs_energy} shows the energy dependence of these mean values for the various track shape observables.  
While some systematic differences between data and Monte Carlo (MC) simulations are observed, the complexity of reproducing the full electron recoil sCMOS image should be acknowledged. Overall, there is satisfactory agreement in track shape, light response, and energy resolution. Future inclusion of effects such as lens-induced distortions and improved saturation modeling could enhance the data–MC agreement.

\newpage
\section{Summary}
\label{sec:summary}
In this paper, a simulation framework for Gaseous Time Projection Chambers with Optical Readout has been developed, focusing on the detailed description of the signal generation and readout for electron recoils. The simulation combines standard particle transport tools in \texttt{Geant4}, for the modeling of the energy deposit in the gas mixture, with a dedicated approach that accounts for ionization electron production and attenuation, avalanche multiplication in the GEM stack (including gain non-uniformities and saturation effects), as well as photon production and sensor background noise. This framework was tuned and validated against data from the LIME prototype of the CYGNO collaboration, which features an active volume of \SI{50}{\litre}, a triple-GEM amplification stage, and an sCMOS camera readout.

The presented simulation provides a realistic modeling of several experimental observables, spanning from integrated light counts and energy resolutions at various source energies to topological variables describing track morphology. In particular, the energy resolutions for \SIrange{3.7}{47}{\keV} X-rays, reconstructed with the same analysis pipeline used for real data, are reproduced to within a few percent. Additionally, key track-shape observables—such as transverse size and amplitude—are reproduced in good agreement with the data. This broad consistency highlights the effectiveness of our modeling of the entire detection chain, from gas diffusion and avalanche formation to optical readout. The resulting realistic detector response is thus a valuable tool for future performance optimization, systematic uncertainty studies, and sensitivity estimates for low-energy signals, including those anticipated in dark matter searches.

\acknowledgments
This project has received funding under the European Union’s Horizon 2020 research and innovation program from the European Research Council (ERC) grant agreement No 818744.
This work is partially supported by ICSC – Centro Nazionale di Ricerca in High Performance Computing, Big Data and Quantum Computing, funded by European Union – NextGenerationEU.

\bibliography{CYGNO}

\end{document}